\documentclass[prl,twocolumn,amsmath,amssymb,groupedaddress]{revtex4}

\usepackage{dcolumn}
\usepackage{graphicx,subfigure}
\usepackage{bm}
\usepackage{verbatim}
\usepackage{amsmath}
\usepackage{amssymb}
\usepackage[T1]{fontenc}
\usepackage{ae,aecompl}
\usepackage{appendix}
\usepackage{float}
\usepackage{color}

\newcommand{\pt}{{\partial}}

\newcommand{\rv}{{\bf r}}

\newcommand{\av}{{\bf a}}

\newcommand{\tv}{{\bm\tau}}
\newcommand{\qv}{{\bf q}}

\newcommand{\oh}{{\frac{1}{2}}}

\newcommand{\grad}{{\bm{\nabla}}}

\newcommand{\be}{\begin{equation}}
\newcommand{\ee}{\end{equation}}
\newcommand{\bea}{\begin{eqnarray}}
\newcommand{\eea}{\end{eqnarray}}
\newcommand{\bse}{\begin{subequations}}
\newcommand{\ese}{\end{subequations}}
\def\rf#1{(\ref{#1})}

\begin{document}
\title{Quantum smectic gauge theory}
\author{Leo Radzihovsky}
\affiliation{
Department of Physics and Center for Theory of Quantum Matter\\
University of Colorado, Boulder, CO 80309}
\date{\today}
\email{radzihov@colorado.edu}

\begin{abstract}
  We present a gauge theory formulation of a two-dimensional quantum
  smectic and its relatives, motivated by their realizations in
  correlated quantum matter. The description gives a unified treatment
  of phonons and topological defects, respectively encoded in a pair
  of coupled gauge fields and corresponding charges. The charges
  exhibit subdimensional constrained quantum dynamics and anomalously
  slow highly anisotropic diffusion of disclinations inside a
  smectic. This approach gives a transparent description of a
  multi-stage quantum melting transition of a two-dimensional
  commensurate crystal (through an incommensurate crystal -- a
  supersolid) into a quantum smectic, that subsequently melts into a
  quantum nematic and isotropic superfluids, all in terms of a
  sequence of Higgs transitions.
\end{abstract}

\maketitle

\noindent{\em Introduction.}
A smectic state of matter, characterized by a uniaxial {\em
  spontaneous} breaking of rotational and translational symmetries is
ubiquitous in classical liquid crystals of highly anisotropic
molecules (e.g., classic 5CB).\cite{ProstDeGennes} Its quantum
realizations range from quantum Hall smectics of a two-dimensional
electron gas at half-filled high Landau
levels\cite{EisensteinQSm,CsathyARCMP,Fogler,Moessner,FradkinKivelsonQHsm,FisherMacdonald,LR_Dorsey},
and ``striped'' spin and charge states of weakly doped correlated
quantum magnets\cite{TranquadaStripes,KivelsonStripes} to the putative
Fulder-Ferrel-Larkin-Ovchinnikov (FFLO) paired superfluids in
imbalanced degenerate atomic gases\cite{LR_VishwanathPRL,LRpra} and
spin-orbit coupled Bose condensates.\cite{LR_ChoiPRL,HuiZhai}

A smectic can emerge from anisotropic partial
melting\cite{HalperinOstlund} of a two-dimensional (2D) crystal,
understood in terms of a Kosterlitz-Thouless (KT)-like,\cite{KT}
single-species dislocation unbinding transition. However, such a 2D
smectic is unstable to thermal fluctuations, driven into a nematic
fluid at any nonzero temperature.\cite{HalperinOstlund,TonerNelson} In
contrast, at zero temperature a 2+1D quantum smectic is a stable state
of matter\cite{LR_VishwanathPRL,LRpra}, whose studies have been
limited to the simplest harmonic-phonons description, typically
neglecting quantum effects of topological defects and of elastic
nonlinearities (as an exception, see e.g.,
Refs.~\onlinecite{LR_VishwanathPRL,LRpra}).

\begin{figure}[htbp]
  \hspace{-0.13in}\includegraphics*[width=0.5\textwidth]{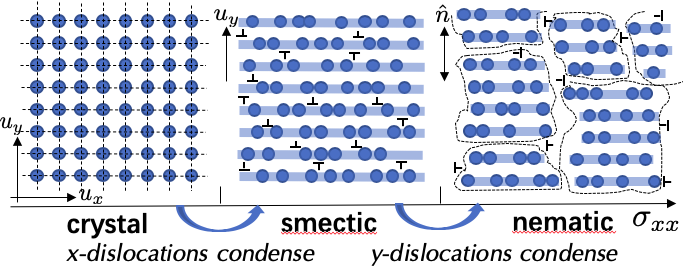}
  \caption{Illustration of quantum melting of a 2D crystal into a
    smectic, followed by smectic-to-nematic melting, respectively
    driven by a condensation of $b_x$ and of $b_y$ dislocations.}
  \label{fig:phase transition}
\end{figure}

Besides intrinsic interest in the smectic and its quantum melting
(freezing) transition into the adjacent quantum nematic (crystal), we
are motivated by a recently discovered duality (previously explored in
other contexts\cite{Zaanen2017}) between elasticity and gapless
``fracton
matter''\cite{PretkoLRdualityPRL2018,PretkoLRsymmetryEnrichedPRL2018,PretkoZhaiLRdualityPRB,KumarPotter19,GromovDualityPRL2019,RadzihovskyHermeleVectorGaugePRL2020}
-- exotic quantum phases that exhibit quasi-particle excitations with
restricted subdimensional mobility. It is natural to explore whether
such restricted mobility survives inside the smectic phase.
\begin{center}
\begin{figure}[htbp]
\centering
 \hspace{0in}\includegraphics*[width=0.48\textwidth]{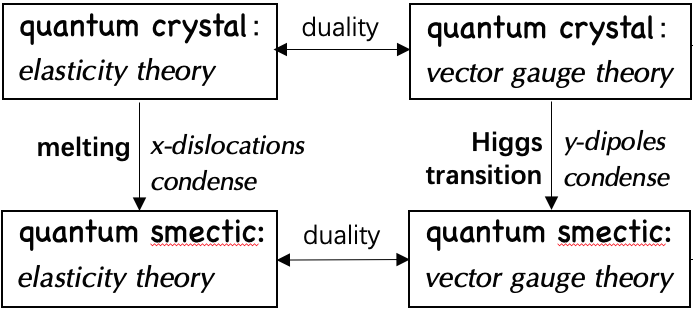}
 \caption{Quantum crystal-smectic duality relations and the associated
   quantum melting transition.}
\label{fig:flowchart}
\end{figure}
\end{center}
\noindent{\em Results.}
In this Letter, we study a time-reversal invariant quantum smectic and
transitions into it via two complementary approaches, summarized in
Fig.\ref{fig:flowchart}.  First, by dualizing a quantum smectic, we
derive a 2+1D coupled U(1) vector gauge theory, with the dual
Hamiltonian density,
\begin{equation}
\begin{split}
  \tilde{\mathcal{H}}_{\text{sm}}=&\frac{1}{2}\kappa{\bf
    E}^2+\frac{1}{2}(\grad \times {\bf A})^2+\frac{1}{2}K
  {\bf e}^2\\
  &+\frac{1}{2}( \grad \times {\av}-\hat{\bf x} \times {\bf
    A})^2-{\bf A} \cdot {\bf J}-{\bf a} \cdot {\bf j},
\end{split}
\label{dualHsm}
\end{equation}
supplemented by the generalized Gauss law constraints,
\begin{eqnarray}
  \grad \cdot {\bf E}&=&p-{\bf e}\cdot \hat{\bf x},\;\;\;
  \grad \cdot {\bf e}=n.
\label{GaussLaws}
\end{eqnarray}
The description is in terms of two coupled $U(1)$ vector gauge fields
with electric fields ${\bf E}$ and ${\bf e}$, and corresponding
canonically conjugate vector potentials ${\bf A}$ and ${\bf a}$,
capturing smectic's gapless phonon degrees of freedom. $p$ and ${\bf
  J}$ are $\hat{\bf x}$-dipole charge and current densities, representing
$\hat{\bf y}$-dislocations, while $n$ and ${\bf j}$ are fractonic charge and
current densities, corresponding to disclinations. The generalized
gauge invariance of \rf{dualHsm} gives coupled continuity equations
for the densities
\begin{subequations}
\begin{eqnarray}
  \partial_t p + \grad \cdot {\bf J}&=&-{\bf j}\cdot\hat{\bf
    x},\label{chargeContinuity_a}\\
  \partial_t n + \grad \cdot {\bf j} &=& 0,\label{chargeContinuity_b}
\end{eqnarray}
    \label{chargeContinuity}
\end{subequations}
where dipole conservation is violated by a nonzero charge current,
${\bf j}\cdot\hat{\bf x}$ along smectic layers.

These equations transparently encode the restricted mobility of the
charges $n$, with their mobility and diffusion coefficient vanishing
along the $\hat{\bf x}$-directed smectic layers, i.e., ${\bf
  j}\cdot\hat{\bf x} = 0$ in the absence of dislocation dipoles. We
thus predict that disclination charges exhibit ``anomalous diffusion''
inside the smectic phase, with anomalous relaxation rate vanishing as
\begin{equation}
\Gamma_k = D k_y^2 + \gamma k_x^4,
\end{equation}
as was also recently explored in
Refs.~\onlinecite{subdiffusionPrinceton,subdiffusionPollmann,Gromov2020}.

Alternatively, we utilize the coupled $U(1)$ vector gauge theory dual
of a 2+1D quantum crystal\cite{RadzihovskyHermeleVectorGaugePRL2020},
``soften'' it into a generalized Abelian-Higgs model, and drive it
through a Higgs transition for one of the dipole species ($\hat{\bf
  y}$-dipoles for coordinate choice in Fig.\ref{fig:phase
  transition}). With the latter a dual description of an anisotropic
quantum melting associated with a condensation of a single type of
dislocations, we thereby obtain a dual gauge theory of a quantum
smectic. As required by consistency, the resulting model is in full
agreement with the first approach of a direct duality of a quantum
smectic, as summarized by the flow chart in Fig.\ref{fig:flowchart}.
The electrostatic limit of the dual gauge theory efficiently
reproduces the model and results of a 2D classical
smectic.\cite{TonerNelson,TonerNSm}

Because a smectic is a condensate of $\hat{\bf x}$-dislocations, it is
necessarily accompanied by a liquid of vacancies and
interstitials. Thus, the dual gauge theory \rf{dualHsm} is implicitly
understood to be coupled (via axion-like ${\cal E}-B$, ${\cal B}-E$
couplings) to a conventional $U(1)$ gauge theory with fields ${\cal E,
  B}$ -- a dual to a liquid of vacancies and
interstitials.\cite{PretkoLRsymmetryEnrichedPRL2018,PretkoZhaiLRdualityPRB}
Since vacancies and interstitials consist of a pair of
oppositely-charged dislocatons, a condensate of the latter necessarily
drives a condensation of the former, requiring a smectic to be an
incommensurate ``super-smectic'' (in contrast to a possibility of two
distinct - Mott insulating ``normal'' (commensurate) and supersolid
(incommensurate)
crystals.\cite{PretkoLRsymmetryEnrichedPRL2018,PretkoZhaiLRdualityPRB,KumarPotter19}).

\noindent{\em Smectic duality.}
We formulate a $2+1$D quantum smectic in terms of a phonon (layer
displacement) ${\bf u} = u(\rv)\hat{\bf y}$ and the unit-normal (layer
orientation) $\hat{\bf n}(\rv)=-\hat{\bf x}\sin\hat\theta + \hat{\bf
  y}\cos\hat\theta \equiv \hat{\bf y} + \delta\hat{\bf n}$ field
operators, and the corresponding canonically conjugate linear and
angular momentum fields, $\pi(\rv)$ and $L(\rv)$, with the Hamiltonian
density,
\begin{equation}
  \mathcal{H}_{\text{sm}}=\frac{1}{2}{\bf \pi}^2
  +\frac{1}{2} L^2+\frac{1}{2}\kappa(\grad u+\delta\hat{\bf
    n})^2+\frac{1}{2} K(\grad \hat{\bf n})^2,
\end{equation}
where $\kappa,K$ are elastic constants.

It is convenient to work with a phase-space path-integral formulation, 
corresponding to the Lagrangian density,
\begin{eqnarray}
  \mathcal{L}_{\text{sm}}&=&\pi \partial_t u+L\partial_t
  \theta-\frac{1}{2}\pi^2-\frac{1}{2}L^2+\frac{1}{2}\kappa^{-1} {\bf
    \sigma}^2+\frac{1}{2} K^{-1} {\tv}^2\nonumber\\
  &-&{\bm\sigma} \cdot \left(\grad u - \hat{\bf x}\theta\right)
  -{\tv} \cdot \grad\theta,
\label{Lsm}
\end{eqnarray}
where we neglected $\theta$ nonlinearities and took the x-axis to be
along the smectic layers. Functionally integrating over the smooth,
single-valued parts of the phonon $u$ and orientation $\theta$ fields,
we obtain coupled constraint equations
\begin{eqnarray}
\partial_t \pi-\grad \cdot {\bm\sigma}=0,\;\;
\partial_t L-\grad \cdot \tv=\hat{\bf x}\cdot{\bm\sigma},
\label{momentumContinuity}
\end{eqnarray}
Newton's laws encoding the linear and angular momenta conservation.

As in electrodynamics, these equations (corresponding to the analog of
Faraday law) are readily solved by expressing densities in terms of
gauge fields,
\begin{eqnarray}
\pi&=&\hat{\bf z} \cdot \left( \grad \times {\bf A}\right),\;\;
{\bm\sigma}=\hat{\bf z}\times(\partial_t{\bf A}+\grad A_0),\\
L&=&\hat{\bf z} \cdot \left(\grad \times {\bf a}-\hat{\bf x} \times {\bf A} \right),\;\;
\tv =\hat{\bf z}\times(\partial_t{\bf a}+\grad a_0 - \hat{\bf x}A_0).
\nonumber
\end{eqnarray}
Reformulating the Lagrangian density \rf{Lsm} in terms of these
Goldstone-mode encoding gauge fields, we obtain the Maxwell part of
the smectic dual Lagrangian,
\begin{eqnarray}
  \mathcal{L}^{\text{sm}}_{\text{M}}&=&\frac{1}{2\kappa} \left( \partial_t {\bf
      A}+\grad A_0\right)^2-\frac{1}{2}\left(\grad \times {\bf A}
  \right)^2\\
  &+&\frac{1}{2K} \left(\partial_t {\bf a}+\grad
    a_0-A_0 \hat{\bf x}\right)^2-\frac{1}{2}\left(\grad \times {\bf
      a}-\hat{\bf x}\times {\bf A}\right)^2,\nonumber
\end{eqnarray}
displaying a nontrivial ``minimal'' coupling between the translational
and orientational gauge fields, which encodes semi-direct product of
spatial translations and rotations. The Lagrangian exhibits a
generalized gauge invariance under transformations,
\begin{subequations}
\begin{eqnarray}
  &&(A_0, {\bf A}) \to A_{\mu}'=\left(A_0-\partial_t\phi, 
    {\bf A}+\grad\phi\right),\\
  &&(a_0,{\bf a}) \to a_{\mu}'=\left(a_0-\partial_t\chi,
    {\bf a}+\grad\chi-\hat{\bf x}\phi\right).\;\;\;\;\;\;
\end{eqnarray}
\label{gauge_transform}
\end{subequations}
The six gauge field degrees of freedom $A_\mu, a_\mu$ reduce to two
physical ones (corresponding to coupled phonon $u$ and orientation
$\theta$ Goldstone modes) after gauge fixing $\phi,\chi$ and
implementing two Gauss law constraints \rf{GaussLaws}.

\begin{center}
\begin{figure}[htbp]
\centering
 \hspace{0in}\includegraphics*[width=0.48\textwidth]{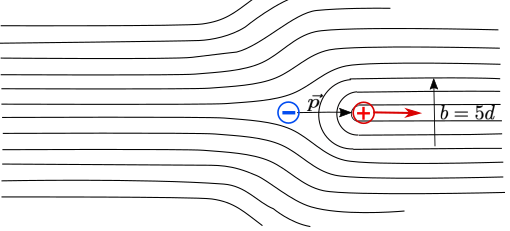}
 \caption{An illustration of restricted along-layers mobility of $+/-$
   disclination (lineon) charges (making up a dislocaton $b$, i.e., a
   dipole $p$) in a quantum smectic, forbidding their separation, that
   corresponds to a nonlocal process of adding a smectic half-layer
   per lattice constant of charge separation.}
\label{fig:mobilityDisclinations}
\end{figure}
\end{center}

To include dislocations and disclinations we allow for the
nonsingle-valued component of $u$ and $\theta$, respectively defined
by
\begin{subequations}
\begin{eqnarray}
  p&=&\hat{\bf z}\cdot\grad\times\grad u,\;\;\;
  {\bf J} =\hat{\bf z}\times\left(\grad\partial_t u
-\partial_t\grad u\right),\\
  n&=&\hat{\bf z}\cdot\grad\times\grad\theta,\;\;\;
  {\bf j}=\hat{\bf z}\times\left(\grad\partial_t \theta
    -\partial_t\grad\theta \right).
\end{eqnarray}
\end{subequations}
This together with $\mathcal{L}^{\text{sm}}_{\text{M}}(A_\mu,a_\mu)$
gives the dual Lagrangian density for the quantum smectic,
\begin{equation}
  \tilde{\mathcal{L}}_{\text{sm}}=\mathcal{L}^{\text{sm}}_{\text{M}}(A_\mu,a_\mu) 
  + {\bf A} \cdot {\bf J} - A_0 p
  + {\bf a} \cdot {\bf j} - a_0 n,
\label{dualLsm}
\end{equation}
corresponding to the Hamiltonian \rf{dualHsm} and Gauss laws
\rf{GaussLaws}. Requiring this dual Lagrangian to be gauge invariant
under \rf{gauge_transform}, immediately leads to coupled continuity
equations \rf{chargeContinuity} for the densities.  The dipole
(dislocation) continuity equation is violated by a nonzero charge
(disclination) current $j_x$ along smectic layers. Thus, in the
absence of gapped dipoles $p$ ($\hat{\bf y}$-dislocations), we find
$j_x=0$, i.e., a motion of isolated lineon charges (disclinations) is
restricted to be transverse to the smectic layers, as moving along the
layers requires $\hat{\bf x}$-dipoles ($\hat{\bf y}$-dislocations, a
nonlocal operation of an insertion of a half-layer of atoms,
illustrated in Fig.\ref{fig:mobilityDisclinations}) that are gapped
inside the smectic ground state.

\noindent{\em Subdiffusive dynamics.}
At finite densities coupled continuity equations \rf{chargeContinuity}
also lead to anomalous subdiffusive dynamics of disclination charges.
To see this, we note that a smectic $\hat{\bf x}$-dipole (a
dislocation $b_y$, illustrated in Fig.\ref{fig:mobilityDisclinations})
corresponds to a separation of charges along smectic layers and so $p
\sim \partial_x n$. The correponding dipole current is controlled by
dipole conservation and is given by a standard Fick's law, ${\bf J} =
-\gamma \grad p$, which translates to ${\bf J}= -\gamma \grad\partial_x
n$, and using first dipole equation gives charge current along smectic
layers to be $j_x = -\grad\cdot{\bf J} = \gamma\nabla^2\partial_x n$. In
contrast there are no constraints on the charge current across the
smectic layers, given by the usual diffusive Ficks form, $j_y =
-D\partial_y n$. Combining these together in the charge continuity
equation \rf{chargeContinuity_b} we obtain
\begin{equation}
  \partial_t n + (\gamma\partial_x^2\nabla^2
  - D\partial_y^2)n = 0, 
\label{anomalousDiffusion}
\end{equation}
displaying subdiffusive dynamics with $i\omega = \Gamma_k\approx D
k_y^2 + \gamma k_x^4$.

\noindent{\em Quantum smectic-to-nematic Higgs transition.}
To access descendant phases and corresponding quantum phase
transitions, we need to treat dislocation and disclination defects in
\rf{dualLsm} as dynamical charges. Following a standard
analysis\cite{DasguptaHalperin,FisherLee,PretkoLRdualityPRL2018} and
focussing on dislocations, we introduce defects core and kinetic
energies (to account for lattice-scale physics) and trace over the
dipole 3-currents $J_\mu = (p, {\bf J})$ to obtain
\begin{eqnarray}
\hspace{-0.2cm}\tilde{\mathcal{L}}_{\text{sm}}&=&
  \frac{g_0}{2}(\partial_t\phi_x - A_0)^2
  + g\cos(\grad\phi_x - {\bf A}) +
  \mathcal{L}^{\text{sm}}_{\text{M}}(A_\mu,a_\mu),
\nonumber\\
  &=&\frac{J}{2}|(\partial_\mu - i A_\mu)\psi_x|^2 - V(|\psi_x|)
  + \mathcal{L}^{\text{sm}}_{\text{M}}(A_\mu,a_\mu).\;\;\;
\label{dualLsmDislocations}
\end{eqnarray}
The Lagrangian is of a relativistic form, encoding dipole neutrality
of a stress-free smectic.  Above, $\phi_x$ is the phase of the
$\hat{\bf x}$-dipole ($\hat{\bf y}$-dislocation) field
$\psi_x=|\psi_x|e^{i\phi_x}$, in the first form we approximated the
resulting Villain potential by its lowest harmonic, and in the second
form went to an equivalent ``soft-spin'' description in terms of
$\psi_x$, with the Landau $U(1)$-invariant potential, $V(|\psi_x|)$.

The $\psi_x=0$ (``non-superconducting'') Coulomb phase of this
generalized Abelian-Higgs model provides a dual description of the
quantum smectic with gauge fields capturing the coupled Goldstone
modes ($u,\theta$) and gapped dual matter $\psi_x$ -- bound
dislocations. Inside the smectic phase the Lagrangian
\rf{dualLsmDislocations} gives an electrostatic interaction $U(\rv)$
between two isolated dipole charges $\psi_x$ with a Fourier
transform $\tilde U(\qv) = \kappa K q_x^2/(\kappa q_y^2 + K
q_x^4)$.\cite{TonerNelson,TonerNSm,LRpra}

The $\psi_x\neq 0$ (``superconducting'' dipole condensate) Higgs
phase, corresponds to a condensed plasma of unbound dislocations, that
gaps out the translational gauge field $A_\mu$, which can therefore be
safely integrated out. This reduces the model to a conventional
Maxwell form for the rotational gauge field $a_\mu$, with
$\mathcal{L}^{\text{nem}}_{\text{M}}(a_\mu)\approx
\mathcal{L}^{\text{sm}}_{\text{M}}(A_\mu \approx 0, a_\mu)=
\frac{1}{2}K^{-1}{\bf e}^2 - \frac{1}{2}(\grad \times {\bf a})^2$,
that is a dual to the quantum xy-model of the nematic,
$\mathcal{L}_{\text{nem}} =\frac{1}{2}(\partial_t\theta)^2 -
\frac{1}{2} K(\grad\theta)^2$. Fluctuation corrections lead to
an anisotropic stiffness and subdominant higher order gradients. As with
the conventional U(1) Higgs (normal-superconductor) transition,
mean-field approximation breaks down for $d+1\leq 4$, and may be
driven first-order by translational gauge-field, $A_\mu$
fluctuations.\cite{HLM} We leave the analysis of the resulting
non-mean-field criticality of the quantum smectic-nematic transition
for a future study.

\noindent{\em Quantum crystal-to-smectic Higgs transition.}
As summarized in Fig.\ref{fig:flowchart}, we can get to the same
smectic gauge dual \rf{dualLsmDislocations} by starting with a gauge
dual of a quantum crystal and taking it through a Higgs transition of
its $\hat{\bf y}$-dipole charges, which corresponds to anisotropic
quantum melting by a condensation of $\hat{\bf x}$-dislocations inside
a crystal. This will generically take place in an explicitly
anisotropic, e.g., $C_2$-symmetric crystal, or by spontaneously
breaking the $C_4$-symmetry of a square lattice, controlled by the
associated Landau potential $V(\psi_x,\psi_y)$. To this end we utilize
the dual U(1) vector gauge theory of a quantum
crystal\cite{RadzihovskyHermeleVectorGaugePRL2020}
\begin{eqnarray}
\hspace{-0.2cm}\tilde{\mathcal{L}}_{\text{cr}}&=&\sum_{k = x,y}
\frac{J_k}{2}|(\partial_\mu - i A^k_\mu)\psi_k|^2 - V(\{\psi_k\})\nonumber\\
&&+ \mathcal{L}^{\text{cr}}_{\text{M}}(A^x_\mu,A^y_\mu,a_\mu),
\label{dualLcrDislocations}
\end{eqnarray}
where $\psi_{k=x,y}$ correspond to $\hat{\bf x}$- and $\hat{\bf
  y}$-oriented dipoles ($\hat{\bf y}$- and $\hat{\bf
  x}$-dislocations), $A^k_\mu,a_\mu$ gauge fields capture the $k =
x,y$ phonons and bond orientational Goldstone modes, $V(\{\psi_k\})$
the Landau potential, and
\begin{eqnarray}
  \mathcal{L}^{\text{cr}}_{\text{M}}&=&
\oh\kappa^{-1}(\partial_t {\bf A}^k + \grad A^k_{0})^2
- \oh(\grad\times {\bf A}^{k})^2\\
&+& \oh K^{-1} (\pt_t a_k + \partial_k a_0 - A^k_{0})^2
- \oh(\grad\times {\bf a} + A_a)^2\nonumber
\end{eqnarray}
is the Maxwell part of the crystal dual Lagrangian, with $A_a =
\epsilon_{ik}A^k_{i}$.\cite{RadzihovskyHermeleVectorGaugePRL2020} The
crystal-smectic partial melting transition corresponds to a
condensation of one of the dipole charges, that according to
Figs.\ref{fig:phase transition},\ref{fig:flowchart} we take to be
$\hat{\bf y}$-dipoles, $\psi_y\neq 0$. This Higgs transition gaps out
$A^y_\mu$, which then can be safely integrated out. To lowest order it
corresponds to $A^y_\mu\approx 0$, reducing the crystal's Maxwell
Lagrangian to that of a smectic \rf{dualLsmDislocations}, with
\begin{eqnarray}
\mathcal{L}^{\text{sm}}_{\text{M}}(A^x_\mu, a_\mu)\approx
\mathcal{L}^{\text{cr}}_{\text{M}}(A^x_\mu,A^y_\mu \approx 0, a_\mu).
\end{eqnarray}

\noindent{\em Vacancies and interstitials.} So far, we have neglected
vacancies and interstitials. As discussed in detail in
Refs.\onlinecite{PretkoLRsymmetryEnrichedPRL2018,PretkoZhaiLRdualityPRB,KumarPotter19},
at zero temperature two qualitatively distinct - commensurate and
incommensurate quantum crystals can appear, respectively distinguished
by Mott insulating and superfluid vacancies and interstitials. In the
former the U(1) symmetry-enriching constraint imposes a glide-only
motion of dislocation, that is broken in the latter, where dipole
dislocation motion is unconstrained. In contrast, we observe that
since a smectic is a condensate of $\hat{\bf x}$-dislocations (created
by $\hat{b}^\dagger_{\bf \hat x}$) and vacancies and interstitials
(created by $a^\dagger$) consist of pairs of oppositely-charged
dislocations (i.e., are disclination charge quadrupoles), the allowed
coupling $\hat{a} \hat{b}^\dagger_{\bf b}\hat{b}^\dagger_{-\bf b}$
drives a condensation of vacancies and interstitions. Thus, a smectic
is necessarily an incommensurate ``super-smectic'', and its dual gauge
theory \rf{dualHsm} is implicitly understood to be coupled (via
axion-like ${\cal E}-B$, ${\cal B}-E$ couplings) to a conventional
$U(1)$ gauge theory with fields ${\cal E, B}$ -- a dual to a liquid of
vacancies and
interstitials.\cite{PretkoLRsymmetryEnrichedPRL2018,PretkoZhaiLRdualityPRB}

\noindent{\em External perturbations.} Repeating the analysis in the
presence of an incommensurate substrate, we find that it reduces the
orientational sector (${\bf e, a}$) from a compact $U(1)$ to $Z_s$
gauge theory (coupled to a noncompact $U(1)$ translational sector
${\bf E, A}$), compactifying the corresponding flux energy
$\frac{1}{2}\left(\grad \times {\bf a}-\hat{\bf x}\times {\bf
    A}\right)^2\rightarrow -\cos\left[2\pi(\grad \times {\bf
    a}-\hat{\bf x}\times {\bf A})/s\right]$, where $s$ is an integer
characterizing orientational commensurability. For $s=1$, the
orientational sector is confined, reducing the model to a conventional
noncompact $U(1)$ gauge theory of the translational sector (${\bf E,
  A}$). Latter can also get confined by a translationally commensurate
substrate, corresponding to the gapping out of the smectic phonon.

A smectic can also be subjected to an external stress. Unlike a
crystal, there is no smectic's resistence to a static shear strain,
$\partial_x u$, as it corresponds to a reorientation of the smectic
order that spontaneously breaks a rotationally invariant fluid. A
compressive stress, $\sigma_{y}$ couples to the strain $\partial_y u$,
that in a gauge theory induces an electric field $E_x$ across the dual
``dielectric''. For $\sigma_y$ above a critical value, a dielectric
breakdown takes place, corresponding to a proliferation of ${\bf\hat
  y}$-dislocations under a super-critical compressive stress. 

\noindent{\em Summary.} Utilizing quantum elasticity
duality\cite{PretkoLRdualityPRL2018,RadzihovskyHermeleVectorGaugePRL2020},
we presented a formulation of a quantum smectic in terms of a
fractonic vector gauge theory. It gives a unified description of
phonons and topological defects and transparently captures the
subdimensional constrained dynamics of the disclination charges,
predicting their subdiffusive hydrodynamics. The dual coupled vector
gauge theory exhibits a unifying global phase diagram of a quantum
crystal (that comes in distinct commensurate and incommensurate -
supersolid forms), smectic, nematic, and isotropic superfluid phases
and describes quantum phase transitions between them in terms of
various Higgs transition. The electrostatic limit of the theory gives
an efficient description and reproduces results of the corresponding
classical phase
transitions.\cite{TonerNelson,TonerNSm,PretkoZhaiLRdualityPRB} We
expect that this dual description will be of value for more detailed
exploration of these phases and corresponding phase transitions.

\emph{Note Added:} After this work was completed we became aware of an
interesting but somewhat orthogonal work by Gromov, and by Gromov and
Moroz, where quantum smectic also
appears.\cite{GromovSm,GromovoMorozSm}

\emph{Acknowledgments}.  I thank Zhengzheng Zhai for help with the
figures and for an earlier collaboration with Michael Pretko, Michael
Hermele.  I acknowledge support by the Simons Investigator Award from
the Simons Foundation, and by the Soft Materials Research Center under
NSF MRSEC Grants DMR-1420736.

\end{document}